\documentclass[%
 reprint,
superscriptaddress,
nofootinbib,
bibnotes,
 amsmath,amssymb,
 aps,
prl,
]{revtex4-2}

\usepackage{graphicx}
\usepackage{dcolumn}
\usepackage{bm}
\usepackage[normalem]{ulem}

\usepackage{amsmath,amsthm, amssymb,epsfig,euscript,array,xcolor,dsfont,comment, physics,graphicx,mathtools,appendix, tensor}

\usepackage[hidelinks]{hyperref}

\newcommand{\V}{\mathcal V}
\newcommand{\Kahler}{K\"ahler }
\newcommand{\lr}[1]{\left(#1\right)}

\renewcommand{\t}{\mathfrak t}
\newcommand{\de}{\partial}

\begin{document}

\title{The gravitational wave landscape of cosmic string networks with varying tension}

\author{Luca Brunelli}
\email{l.brunelli@unibo.it}
\affiliation{Dipartimento di Fisica e Astronomia, Universit\`a di Bologna, via Irnerio 46, 40126 Bologna, Italy} 
\affiliation{INFN, Sezione di Bologna, viale Berti Pichat 6/2, 40127 Bologna, Italy 
}

\author{Filippo Revello}
\email{filippo.revello@kuleuven.be}
\affiliation{Instituut voor Theoretische Fysica, KU Leuven, Celestijnenlaan 200D, B-3001 Leuven, Belgium}
\affiliation{Leuven Gravity Institute, KU Leuven, Celestijnenlaan 200D, B-2415 Leuven, Belgium}

\author{Gonzalo Villa}
\email{gv297@cam.ac.uk}
\affiliation{DAMTP, Centre for Mathematical Sciences, University of Cambridge, \\ Wilberforce Road, Cambridge, CB3 0WA, UK}

\begin{abstract}
We fully classify the phenomenology of gravitational wave emission from scaling cosmic string networks with varying tension and compute the spectral indices of the resulting stochastic backgrounds. In string compactifications, periods of varying tension occur when moduli acquire a time-dependence. We present concrete examples in type IIB string theory as D3- and NS5- branes wrapping internal cycles, which become dynamical due to the effect of moduli potentials. Moreover, we use Swampland constraints to derive general bounds on the allowed time-variation of the effective string tension in FLRW backgrounds and on the resulting spectral indices.\end{abstract}

\maketitle

\section{Introduction}
Since the discovery that D-branes are fundamental, dynamical components of string theory, the study of extended heavy objects has become a central topic in the subject.
As any other physical quantity in string theory, the properties of these objects (in Planck units) are set dynamically in a given vacuum, and so in particular they depend on the expectation values of moduli fields (and other quantities like gauge fluxes).
Whilst in highly supersymmetric vacua these moduli remain flat directions, in any realistic string vacuum supersymmetry ought to be broken at some high energy scale.
Among other things, supersymmetry breaking leads to isolated points in the moduli space becoming energetically preferred: the main target of string phenomenology is to identify which of these vacua resembles our Universe.\footnote{While remarkable progress in this direction has been made, it is fair to say that no concrete construction has all ingredients required to claim victory (a fully controlled construction of metastable de Sitter vacua in particular still remains an open question, see \cite{McAllister:2023vgy,McAllister:2024lnt} for state of the art progress in this direction).}

A related question is what are the dynamics that lead to this vacuum after, say, a period of inflation.
Remarkably, one can make general statements about this question without detailed knowledge of the vacuum itself.
More concretely, this paper studies the dynamics of light moduli rolling down exponential potentials, both being consequences of accidental scaling symmetries, a robust feature of string vacua~\cite{Burgess:2020qsc}.

A particularly interesting feature of these exponential potentials is that, in a cosmological background, the fields reach scaling configurations, where their energy density redshifts in the same way as that of the background.
Moreover, such epochs involve an order one of Planckian displacements per e-fold \cite{Wetterich:1987fm,Ferreira:1997hj,Copeland:1997et}, so that physical scales (set by the moduli) can vary substantially with time.\footnote{K\"{a}hler moduli can be displaced from their late-time minimum due to both quantum fluctuations during inflation and their generic couplings to other sources of energy.}
In phenomenologically relevant scenarios this dynamics ends with the scalars reaching the minima of a more complicated potential for which the exponential is a good approximation in a particular regime of field space.
Recent discussions~\cite{Conlon:2022pnx,Apers:2022cyl,Revello:2023hro,Apers:2024ffe, Conlon:2024uob,Revello:2024gwa, Brunelli:2025ems, Ghoshal:2025tlk, SanchezGonzalez:2025uco, Brunelli:2025eif, Conlon:2025mqt, Chun:2025ret} have renewed attention on the effect of rolling moduli after inflation and their effects on cosmic string dynamics, see \cite{Cicoli:2023opf} for a review of string cosmology.

Interestingly, cosmic strings stretching further than the cosmic horizon represent another system that can reach a cosmological scaling regime. Such scaling configurations are achieved through a balance of the energy that enters the horizon as the universe expands, and the energy loss due to intercommutation of these objects, which generate unstable subhorizon loops.
The latter generically release gravitational waves (GWs) and, depending on the microphysics that generates them, perhaps other massless modes (\emph{e.g.} axions~\cite{Davis:1986xc}).
Due to the scaling regime, the amplitude of the resulting gravitational wave background today can be sizeable throughout many orders of magnitude in frequency (each correlated with a given time), with its spectral index depending on the equation of state of the background.
Gravitational waves from cosmic superstring networks are a well studied subject and are active targets of current and upcoming gravitational wave experiments \cite{Ellis:2023tsl,Avgoustidis:2025svu}.\footnote{Our focus is strictly on the spectral indices that are featured in the GW spectrum, which can be calculated in a model-agnostic way. Model-dependent assumptions are required in order to make concrete predictions for overall amplitudes and frequency ranges where such spectral indices occur. An analysis of the detection prospects in current and future GW detectors is therefore beyond the scope of the paper.
More concrete scenarios have been discussed in Ref.~\cite{Ghoshal:2025tlk}.}

As a consequence of scaling, if present throughout the universe, gravitational radiation from cosmic strings could give rise to observational access to epochs before Big Bang Nucleosynthesis (BBN) \cite{Allahverdi:2020bys}.
Since the properties of these objects are given by the moduli, moduli dynamics in the early universe alters the gravitational wave spectrum.
Inspired by the Swampland program, we give general bounds on the rate of variation of the tension, $\mu(t)$, which in turn bound the possible spectral indices provided a scaling regime is achieved.
An analytic study providing evidence that this is the case when the tension varies is given in~\cite{Revello:2024gwa}, as long as the possibly growing loops do not have time to percolate~\cite{Conlon:2024uob}.

Current quantitative understanding of how moduli-induced variations of the tension translate into a GW spectrum is restricted to the dependence on the overall volume modulus for the case of fundamental strings~\cite{Ghoshal:2025tlk}.
In this paper, we explore a qualitatively different system in the context of type IIB string theory: $p$-branes wrapped around $(p-1)$-cycles of an internal manifold whose volume is fixed and admits a product form.\footnote{The latter assumption is important, because it gives rise to exponential potentials.}
The tension of these effective strings depends on the volume of the wrapped cycle and we show that the dynamics of this moduli, assuming the conjecture in \cite{Berg:2007wt}, evolves as to prefer scalings $\mu \sim t^{\pm1}, \, \mu \sim t^{\pm 1/2}$. The case with $\mu \sim t^{-1}$ (for wrapped D3 branes) displays a spectral index which is independent of the equation of state of the background.

Lastly, the GW phenomenology of scaling cosmic string networks with varying tension has not been fully explored in the literature (for early work on cosmic strings with varying tension, see \cite{Emond:2021vts}).
Ref.~\cite{Ghoshal:2025tlk} studied the GW spectrum for particular values of the equation of state of the background and the rate of variation of the tension.
It was shown that in these cases the phenomenology is different to the constant tension case in that the largest contribution to the GW spectrum at a given frequency is sourced by loops that emit in the relevant frequency bin as soon as they are created.
This is in contrast to the case with constant tension, where the loops that source the maximum power in GWs do so after radiating half of their energy.
In this paper we fully classify the possibilities, identifying three different scenarios.
The results are shown in Fig.~\ref{fig:pn}, and we compute the associated spectral index in all regions.
As an application of our Swampland-inspired bounds, we exclude a region of parameter space where the rate of variation of the tension is too large.

\section{Time-dependent string tension}
\subsection{A Swampland Argument}
Borrowing some ideas from the Swampland program, we provide a succinct argument to constrain the time evolution of the string tension in FLRW cosmologies. For both fundamental and field theory strings, the latter can be viewed as the square of some effective field theory (EFT) cutoff. This is clear in the case of fundamental strings, where the tension is related to the string scale $m_s$ as $\mu \sim m_s^2$. As another example, for axion strings $\mu \sim 2 \pi S_{\rm inst} f_a^2$, where $S_{\rm inst}$ is the relevant instanton action and the axion decay constant $f_a$ may also be thought of as a cutoff \cite{Reece:2025thc}.
For strings which are fundamental in nature (\emph{i.e.} either F-strings or wrapped D-branes), it is natural to expect $\mu \geq \Lambda_{s}^2$, where $\Lambda_{s}$ is an upper bound on the quantum-gravitational cutoff known as the species scale \cite{Dvali:2007hz,Dvali:2007wp}.
As discussed in \cite{Martucci:2024trp,Grieco:2025bjy} for the case of EFT strings \cite{Lanza:2020qmt,Lanza:2021udy}, fundamental strings cannot be resolved within the EFT regime, and therefore the mass of their would-be quantised excitations should lie above $\Lambda_s$. 

According to the Emergent String Conjecture (ESC) \cite{Lee:2019wij}, any asymptotic limit in moduli space is realised either as a decompactification limit or a tension-less string.
In that case, $\Lambda_s$ can be identified with the higher-dimensional Planck scale or the string scale respectively. Inspired by the ESC, in \cite{vandeHeisteeg:2023ubh,vandeHeisteeg:2023dlw} it was suggested that (see also \cite{Castellano:2023stg}) the gradient of the species scale be bounded as
\begin{equation}\label{eq:dLambda}
    \Bigg \lvert \frac{\nabla_{\phi} \Lambda_{s}}{\Lambda_{s}}\Bigg \rvert^2 \leq \frac{1}{M_{P,d}^{d-2}} \frac{1}{d-2},
\end{equation}
across the whole of moduli space (including the interior). In the above, the norm of the gradient is taken with respect the field space metric $G_{ij}$, and $d$ is the number of spacetime dimensions of the EFT. If we assume a system of $N$ spatially homogeneous scalars $\phi_i$,  coupled to arbitrary fluids and with a non-zero scalar potential $V(\phi^i)$, the first Friedmann equation reads
\begin{equation}\label{eq:friedman}
    \alpha_d\,  H^2 M_{P,d}^{d-2} = \frac{1}{2}G_{i j}(\phi) \frac{{\rm d} \phi^i}{ { \rm d} t} \frac{{\rm d} \phi^j}{ { \rm d}t} +V(\phi^i)+ \sum_j \rho_j(\phi^i),
\end{equation}
with $\alpha_d=(d-1)(d-2)/2$.
If $V(\phi_i), \, \rho_i >0$ we can combine the two to write
\begin{equation}\label{eq:bound}
     \bigg \lvert \frac{1}{\Lambda_{s}} \frac{{\rm d} \Lambda_{s}}{ { \rm d }t }\bigg \rvert \leq \sqrt{ \bigg \lvert \frac{\nabla_{\phi} \Lambda_{s}}{\Lambda_{s}}\bigg \rvert^2\, 
 \bigg \lvert \frac{{\rm d} \phi^i}{ { \rm d} t}   \bigg \rvert ^2}  \leq \sqrt{d-1} \, H,
\end{equation}
where $ \big \lvert \frac{{\rm d} \phi^i}{ { \rm d} t} \big \rvert^2 = G_{ij} \frac{{\rm d} \phi^i}{ { \rm d} t} \frac{{\rm d} \phi^j}{ { \rm d} t} $ and we have applied the Cauchy-Schwarz inequality to the two vectors $\sqrt{G}_{i j}\partial_{\mu} \phi^i$ and $\sqrt{G}^{i j} \frac{\partial \Lambda_{s}}{\partial \phi^i}$ ($G$ is positive definite). Notice how the bound \eqref{eq:bound} is invariant under time reparametrisation, and hence valid in any frame. A first application is that, if $\Lambda_{s}$ is monotonically decreasing with time (as one expects along paths approaching the boundary of moduli space), \eqref{eq:bound} can be integrated to give
\begin{equation}\label{eq:lambdabound}
    \Lambda_{s}(t) \geq \Lambda_{s}(t_0) \left(\frac{a(t_0)}{a(t)} \right)^{\sqrt{d-1}}.
\end{equation}
This condition is weaker than (and implied by) the requirement $\Lambda_{s} \gtrsim \Lambda_{s}(t_0) \, (a(t_0)/ a(t))$, necessary for the consistency of the EFT at late times \cite{Bedroya:2025ris}. Moreover, since $H \ll \Lambda_s$ in order for the EFT picture to be valid, it implies
\begin{equation}\label{eq:pp}
     \bigg \lvert \frac{1}{\Lambda_{s}^2} \frac{{\rm d} \Lambda_{s}}{ {\rm d} t}\bigg \rvert \leq \sqrt{d-1} \frac{H}{\Lambda_{s}} \ll 1,
\end{equation}
implying, among other things, that gravitational production of heavy states with mass $m \gtrsim \Lambda_{s}$ is suppressed \cite{Baumann:2014nda}. From the reasonable assumption that $\mu \geq \Lambda_s^2$, we can then turn \eqref{eq:lambdabound} into a bound on the time-dependence of string tension:
\begin{equation}\label{eq:mubound}
  \mu(t)  \geq \mu(t_0) \left(\frac{a(t_0)}{a(t)} \right)^{2\sqrt{d-1}}.
\end{equation}
For a cosmology that is dominated by a fluid with $\rho \sim a^{-n}$, and if $\mu \sim t^{-q}$,\footnote{Close to a boundary of moduli space, as the string tension will depend exponentially on scalar fields, and the latter typically evolve logarithmically with time.} Eq. \eqref{eq:mubound} translates to
\begin{equation}\label{eq:b1}
    qn \leq 4 \sqrt{d-1}.
\end{equation}
The latter applies to any scenario with $q \geq 0$.
A decreasing tension is particularly interesting for our purposes because it induces growth in the GW spectrum at high frequencies.

\subsection{Decompactification \& tensionless limits}

To obtain some intuition on the above, let us now consider some simple examples arising in (toy models of) compactifications of 10d string theory/supergravity to 4d.
We first discuss the cases in which the inequality~\eqref{eq:mubound} is saturated and then provide more constraining bounds on the rate of variation of the tension provided the dilaton is fixed.

The inequality~\eqref{eq:mubound} is saturated when Eq.~\eqref{eq:friedman} is dominated by the kinetic energy of the field (\emph{i.e.} kination), with this rolling along the direction where \eqref{eq:dLambda} also holds with an equal sign.
In particular, the latter happens in emergent string limits, where $ \Lambda_{s} \sim m_s$.

Let us discuss explicit examples in a given field space.
Without warping, one can express the Einstein frame volume $\mathcal{V}$ (in units of $m_s$) and string coupling $g_s$ as:
\begin{equation}
    \mathcal{V} \sim e^{\sqrt{\frac{3}{2}} \chi} \quad \quad g_s \sim e^{-\sqrt{2} \varphi},
\end{equation}
where $\chi$ and $\varphi$ are the (4d) canonically normalised volume and dilaton respectively and we have set the 4d Planck mass $M_{P,4}=1$.

For fundamental strings
\begin{equation}
    \mu_{\rm F1} \sim  \frac{g_s^{1/2}}{\mathcal{V}} \sim e^{-\frac{1}{\sqrt{2}} \varphi -\sqrt{\frac{3}{2}} \chi},
\end{equation}
and $|\nabla \mu_{\rm F1}|/ \mu_{\rm F1} = \sqrt{2}$, realising the upper bound in Eq. \eqref{eq:dLambda}.\footnote{In fact, this is true in any number of dimensions, as can be inferred from the species-scale vectors of \cite{Castellano:2023stg}.}
Then, Eq.~\eqref{eq:bound} can be saturated by kination of the ``4d dilaton", \emph{i.e.} the linear combination of $\chi$ and $\phi$ parallel to $\nabla_i \mu_{\rm F1}$. As another example, emergent string limits in 4d can also arise from NS5 branes wrapping 4-cycles \cite{Lee:2019wij}.
Anticipating part of the next section, let us consider compactification on a Calabi-Yau (CY) space with a fibred structure, such that the volume can be written in terms of $4$-cycle volumes $\tau_1,\tau_2$ as $\mathcal{V} = \alpha \, \tau_2 \sqrt{\tau_1}$, for some constant $\alpha$. In terms of the canonically normalised scalars $\chi$ and $\chi_1$,
\begin{equation}
    \mathcal{V}= e^{\sqrt{\frac{3}{2}} \chi} \quad \quad  \tau_1 = e^{\sqrt{\frac{2}{3}} \chi+\frac{2}{\sqrt{3}} \chi_1},
\end{equation}
and the tension of an NS5 wrapped around $\tau_1$ is
\begin{equation}
    \mu_{{\rm NS5}} \sim \frac{m_s^2 \tau_1}{g_s^2} \sim e^{+\frac{1}{\sqrt{2}} \varphi-\left(\sqrt{\frac{3}{2}} -\sqrt{\frac{2}{3}}\right) \chi+\frac{2}{\sqrt{3}} \chi_1}.
\end{equation}
Again, $|\nabla \mu_{\rm NS5}|/ \mu_{\rm NS5} = \sqrt{2}$, precisely along the direction where $\mu_{\rm F1}$ remains constant. Notice how  both of these cases feature the growth region of \cite{Brunelli:2025ems} in the respective phase space, and provide new examples of string loops that can grow in comoving coordinates \cite{Conlon:2024uob,SanchezGonzalez:2025uco,Brunelli:2025eif}.

From a phenomenological perspective, the previous

examples would be relevant for compactifications where the dilaton is light \cite{Chauhan:2025rdj} and can be cosmologically active. However, let us also notice how a large field excursion for $\varphi$ would likely drive the universe to a vacuum where $g_s \ll 1$, presenting challenges for realistic Standard Model embeddings with $\alpha_{SM} \sim 10^{-2}$. For this reason, it is interesting to examine how much the bound \eqref{eq:bound} can be tightened if one assumes a constant string coupling. From the ESC, the only other option is given by a decompactification limit from $d+k \leq 10$ to $d$ dimensions.
In that case, the analogue of Eq.~\eqref{eq:dLambda} is (reintroducing $M_{P,d}$):
\begin{equation}
    M_{P,d}^{d-2}\Bigg \lvert \frac{\nabla_{\phi} \Lambda_{s}}{\Lambda_{s}}\Bigg \rvert^2 \leq 
    \frac{k}{(d+k-2)(d-2)} \leq 
    \frac{10-d}{8(d-2)},
\end{equation}
which leads to the stronger bound:
\begin{equation}\label{eq:b2}
    q  n \leq 2 \sqrt{\frac{(10-d)(d-1)}{2}}.
\end{equation}
In $d=4$, for example, the above is saturated for F1 strings and volume kination, giving $q n =6$ \cite{Revello:2024gwa}.

The above discussion provides bounds which are saturated along particular directions in field space.
In actual models without supersymmetry, the induced scalar potentials will determine preferred directions for the dynamics to occur.
In the next sections, we study such potentials and the dynamics they induce.

\subsection{Scaling solutions}
The bound \eqref{eq:bound} is valid at any time and for an arbitrary trajectory $\left\{ \phi_i(t) \right\}$, including transients between different regimes or oscillatory solutions. On the other hand, for the purpose of GW emission we are interested in solutions lasting for a significant number of e-folds, so that the string network has enough time to form and achieve scaling. For concreteness, let us specialise to the case where $d=4$, $G_{ij} = \delta_{ij}$ and there is a single additional fluid with equation of state $P_{\gamma}= (\gamma-1) \rho_{\gamma}$, usually taken to be matter ($\gamma= 1$) or radiation ($\gamma=\frac{4}{3}$). In that case, one of the Friedmann equations can be written as
\begin{equation}\label{eq:fried}
    \dot{H}= - \frac{1}{2 M_P^2} \left[ \sum_i \dot{\phi}_i^2+ \gamma \rho_{\gamma}\right].
\end{equation}
If the potential $V(\phi_i)$ takes a relatively simple form - sums and products of exponentials in the $\phi_i$ - the full equations of motion can be rewritten as an autonomous dynamical system (see \emph{e.g.} \cite{Hartong:2006rt,Shiu:2025ycw} for the general case including a fluid). Fixed points of the dynamical system are known as \emph{scaling solutions}, and are characterised by constant ratios between the (kinetic and potential) energy densities of each scalar field, as well as that of additional components. Notice that we are not necessarily interested in stable fixed points - unstable or saddle points can also give rise to phases lasting many e-folds for sufficiently fine-tuned initial conditions (which must be justified on a physical basis). An example of this is kination, which is always unstable in the presence of matter/radiation, but can last a significant number of e-folds under certain conditions~\cite{Conlon:2022pnx,Apers:2024ffe}. In this setting, scaling solutions are always characterised by a power-law expression for the scale factor, $a(t) \sim t^{2/n}$ (with $n \leq 6$). Then, \eqref{eq:fried} implies ($M_P=1)$:
\begin{equation}\label{eq:bt}
    \bigg \lvert \frac{{\rm d} \phi^i}{ { \rm d} t}   \bigg \rvert \leq \sqrt{\sum_i \left(\frac{{\rm d} \phi^i}{ { \rm d} t}^2 \right)}  \leq \sqrt{ n-3\gamma \Omega }\, H,
\end{equation}
where $\Omega = \frac{\rho_{\gamma}}{3H^2}$ is constant along the solution. The above equation can be used to derive a tighter version of the bound \eqref{eq:bound} in the case of scaling solutions. In this particular case, the analogues of \eqref{eq:b1} and \eqref{eq:b2} become
\begin{equation}\label{eq:b3}
   q n \leq 2 \sqrt{2n}, \quad \quad \quad q n \leq  \sqrt{6n},
\end{equation}
for general and decompactification limits respectively (in $d=4$). The bound is saturated when $\Omega=0$ in \eqref{eq:bt}, that is scaling solutions dominated by the scalar fields.
Such solutions,
however, can only arise for shallow enough scalar potentials, and are
often unstable. In the presence of a fluid, \emph{tracker} solutions with $\Omega \neq 0$ are typically favoured by energy considerations \cite{Shiu:2024sbe}.

To illustrate this last point, let us quickly recall what happens in the simplest situation involving a single scalar field with a potential $V=V_0 e^{-\lambda \phi}$ in $d=4$ \cite{Copeland:1997et}. For $\lambda^2 <6$, there exists a scaling fixed point with $\Omega =0$ and $n= \lambda^2$, where the time-derivative of the field is 
\begin{equation}
    \dot{\phi}= \lambda H =\sqrt{n} H.
\end{equation} 
This saturates the bound \eqref{eq:bt}, and the corresponding solution is stable only for $\lambda^2 < 3 \gamma$.
Instead, for $\lambda^2 > 3 \gamma$, there exists an attractor where the field tracks the energy density of the fluid ($n= 3 \gamma$) and
\begin{equation}\label{eq:phitr}
 \dot{\phi}= \frac{3 \gamma}{\lambda}H < \sqrt{n} H.  
\end{equation}
We will discuss these type of solutions in what follows.

\section{Novel scenarios in Type IIB}

In this section, we discuss a few explicit examples of time-dependent tension in simplified string theory compactifications with broken supersymmetry.
In order to be fully explicit, it is most convenient to work in a framework where moduli potentials are well understood.
As emphasized in the previous section, moduli potentials will invariably affect their cosmological dynamics, and are therefore a crucial ingredient to derive the exact time-dependence of the tension.\footnote{As opposed to the bounds assuming the largest possible $|\dot{\phi}|$, as in kination.}
For this reason, we specialise to type IIB flux compactifications: in particular the LVS scenario~\cite{Balasubramanian:2005zx} and variations thereof.
From the underlying GKP construction~\cite{Giddings:2001yu}, complex structure moduli (including the dilaton) are stabilised at parametrically heavy scales, and do not play a dynamical role. On the other hand, the \Kahler moduli are flat directions at tree-level (thanks to the no-scale property), and only obtain a potential from perturbative and non-perturbative corrections.
We thus restrict our attention to the subset of moduli space spanned by \Kahler moduli, and identify the relevant directions along which the dynamics occurs.

In this setting, together with F1s and D1s, the only objects that can be sufficiently stable to form a network of effective strings are D3, D7 and NS5 branes~\cite{Copeland:2003bj}. The dependence of truly one-dimensional objects on the \Kahler moduli arises through the overall volume modulus, whilst the tension of D7-strings does not depend on these at all. The dynamics of this modulus and its implications for the tensions of fundamental strings has already been studied extensively in \cite{Revello:2024gwa,Conlon:2024uob,Ghoshal:2025tlk,SanchezGonzalez:2025uco} and, since the dependence of the tension of D1-branes on K\"{a}hler moduli is the same, we will not consider either case here.
In the following, we will focus on the remaining cases:
effective strings arising from either a D3-brane wrapping a 2-cycle (or an NS5 wrapping the dual 4-cycle) in a manifold with a factored volume, expanding on what was done in \cite{Brunelli:2025ems}.
Such strings will also turn out to have very interesting properties in terms of their gravitational wave emission.

\subsection{Manifolds with Factored Volume}

We will consider either a toroidal manifold  or a fibred manifold with volumes respectively: 
\begin{equation}\label{eq:volume t_i}
    \V_{\rm tor} = \t_1\t_2 \t_3 \, , \qquad \V_{\rm fib}=\t_1\t_2^2 \,,  
\end{equation}
where the $\t_i$'s are volumes of 2-cycles in string units.

Inverting the relation between corresponding 2- and 4-cycles, namely $\tau_i =  \de \V/\de \t_i$, one gets, up to overall constants, the volume in terms of the 4-cycle volumes:
\begin{equation}\label{eq:volume tau}
    \V_{\rm tor}= \sqrt{\tau_1 \tau_2 \tau_3} \, , \qquad \V_{\rm fib}= \tau_2 \sqrt{\tau_1} \, .
\end{equation}

The tree-level \Kahler potential $K \equiv -2\ln (\V)$ for the \Kahler moduli will be, up to a constant term:
\begin{equation}\label{eq:kahler}
    K_{\rm tor}=-\sum_{i=1}^{3}\ln (\tau_i) \, , \qquad K_{\rm fib}=-2\ln(\tau_2)-\ln(\tau_1)\, .
\end{equation}
and displays a no-scale structure. 

\subsection{Contributions to the Scalar Potential}

A well known consequence of the tree-level scale invariance of string compactifications is that K\"{a}hler moduli span flat directions with zero vacuum energy at tree level.
In K\"{a}hler moduli stabilisation schemes of type IIB string theory, the overall volume acquires a potential due to a combination of perturbative and non-perturbative corrections (see Ref.~\cite{McAllister:2023vgy} for a review).
Generically, however, the compactification manifolds have $N$ 4-cycles and so the overall stabilisation mechanism leaves $N-1$~\cite{Kachru:2003aw} or $N-2$~\cite{Balasubramanian:2005zx} flat directions at this level of approximation. 
The dynamics and stabilisation of these additional directions due to further corrections has been studied in many contexts, with phenomenological applications from inflation to dark energy (see~\cite{Cicoli:2023opf} and references therein).
In the present context, the dynamics induced in the remaining moduli gives rise to a particular time-dependence for the tension of effective strings obtained by D3-branes wrapping one of the 2-cycles in \eqref{eq:volume t_i}, or NS5-branes wrapping the dual cycles. 

In \cite{Berg:2005ja}, the authors studied string-loop corrections to the \Kahler potential in a toroidal orientifold.
They found that the only non-vanishing $\mathcal N = 1$ contributions arise from KK mode exchange between parallel stacks of branes (KK-loops), or winding corrections at the intersections of stacks (W-loops).
The Berg-Haack-Pajer (BHP) conjecture~\cite{Berg:2007wt} proposes a generalisation of these results for the form of loop corrections to the \Kahler potential in the case of a Calabi-Yau orientifold. Due to an extended no-scale structure found in \cite{Cicoli:2008gp}, the leading KK correction takes the form $\delta V_{KK}(\tau_i) \propto  K_{\rm tree}^{ii}$, with no summation over $i$.
For a product-form volume (cf.~\eqref{eq:volume tau}), this translates into corrections to the scalar potential of the form:
\begin{equation}
    \delta V_{\rm KK} (\tau_i) = \frac{C_i}{\V^2} \, \frac{1}{\tau_i^2}\label{eq:KK potential} \, ,
\end{equation}
where $C_i$ are functions of complex structure moduli and the string coupling, which we take as constants.
Low-energy arguments supporting these scalings were provided in~\cite{Cicoli:2007xp} and more recently in~\cite{Gao:2022uop, Bansal:2024uzr}, where it was argued that these arise in the absence of branes as well.  
In what follows we assume that the overall volume is fixed and that these are the leading corrections to the potential (as is known to be true in concrete examples~\cite{Cicoli:2008gp}).
This allows us to study the dynamics induced on (a subset of) the remaining moduli.

\subsection{The fibre}

The strategy is to identify the orthogonal direction to the volume which the potential in Eq.~\eqref{eq:KK potential} renders dynamical.
In a fibred CY, we can rewrite the moduli in terms of two orthogonal directions in field space: the overall volume $\V\propto \tau_2 \sqrt{\tau_1}$ and $u  \propto \tau_1/\tau_2$.
We note that the assumption that the volume is fixed at a higher mass than the field $u$ is well justified for perturbative moduli stabilisation of fibred manifolds \cite{Cicoli:2008gp}.
From the \Kahler potential \eqref{eq:kahler}, one can work out the canonical normalisation of the fields:
\begin{equation}
    \V \sim e^{\sqrt{\frac{3}{2}} \chi}, \quad u \sim e^{\sqrt{3} \chi_1}.
\end{equation}
This leads to the following scalings of the fibre 2/4-cycle with $\chi_1$:
\begin{equation}
    \t_2 \sim \V^{1/3} e^{\frac{1}{\sqrt{3}}\chi_1}, \quad \tau_1 \sim \V^{2/3} e^{\frac{2}{\sqrt{3}}\chi_1}
\end{equation}
and that of the base 2/4-cycles:
\begin{equation}\label{eq:t1-T2}
    \t_1 \sim \V^{1/3} e^{-\frac{2}{\sqrt{3}}\chi_1}, \quad \tau_2 \sim \V^{2/3} e^{-\frac{1}{\sqrt{3}}\chi_1}.
\end{equation}
We can see that, for a fibred CY, and assuming $\tau_2 \gg \tau_1$, the leading order KK correction to the scalar potential for the fibre is:
\begin{equation}\label{eq:fiber-potential}
    \delta V_{\rm KK}(\tau_1) \sim \frac{C_1}{\V^2} \frac{1}{\tau_1^2}\sim  e^{-\frac{4}{\sqrt{3}}\chi_1}   \, , 
\end{equation}
where the last term focuses on $\chi_1$ dependence.
This term dominates in the scalar potential to the left of the minimum.  
In what follows we assume that the initial conditions are such that the field becomes dynamical somewhere in this region of field space, before settling down into its minimum. 

Exponential potentials as in Eq.~\eqref{eq:fiber-potential} are generic in string compactifications and have been thoroughly studied (see~\cite{Apers:2024ffe} and references therein for recent discussions). For cosmological purposes, we assume the existence of a perfect fluid contributing to the energy density of the Universe, inducing Hubble friction to the dynamics of $\chi_1$.
Fixed points of the dynamics then correspond to the scaling solutions of the previous section. According to our analysis, for this particular value of the exponent $\lambda$ (we recall $V=V_0 \, e^{-\lambda \phi}$), both field-dominated and fluid (co)-dominated scaling solutions exist, although only the latter are an attractor for $\gamma \geq 1$~\cite{Copeland:1997et}.
While the field-dominated solution could in principle also last for several e-folds, we do not have a physical motivation to justify the necessary (fine-tuned) initial conditions.\footnote{Volume modulus kination after inflation is however an example where these initial conditions are well motivated ~\cite{Conlon:2022pnx}.
}
Therefore, from now on we will only consider the attractor solutions, given by a fluid tracker. From \eqref{eq:phitr}, the evolution along the attractor is described by\footnote{
The parameter $\chi_1 (t_0)$ is determined by $V_0$, $\lambda$ and the initial value of the potential $V(t_0)$.}
\begin{equation}\label{eq:sol-scale}
    \chi_1 (t) =\chi_{1} (t_0)+ \frac{2}{\lambda}\ln \left(t/t_0 \right) \, ,
\end{equation}
where the dependence on $\gamma$ has dropped out.
According to Eqs.~\eqref{eq:t1-T2} and~\eqref{eq:sol-scale}, this implies that the time-dependence of the tension of a three-dimensional brane wrapping $\t_1$ (which is proportional to $\t_1$) is:
\begin{equation}\label{eq:tension-variation}
    \mu(t)=\mu_0 \left( \frac{t_0}{t}\right) \, .
\end{equation}
The opposite scaling $\mu \sim t $ is found for a five-brane wrapping $\tau_1$.
Similarly, the tension of a three-brane wrapping $\t_2$ grows:
\begin{equation}
    \mu(t)=\mu_0 \left(\frac{t}{t_0}\right)^{1/2} \, ,
\end{equation}
while the tension of a five-brane wrapping the dual cycle $\tau_2$ decreases like $t^{-1/2}$.

The regime $\tau_1 \gg \tau_2$ does not admit a similar analysis because $\lambda=-2/\sqrt{3}$, and so $\lambda^2=4/3<n$ for radiation ($n=4$) and matter ($n=3$), so the tracker is unstable.
For these values of the parameters there also exist stable, field-dominated solutions. The latter would correspond to accelerated expansion and are not phenomenologically relevant for our purposes.\footnote{Note that this is not in contradiction with the strong de Sitter conjecture \cite{Rudelius:2021oaz}, which bounds $|\nabla V/V|< \sqrt{2} \, M_p$ in the asymptotics of moduli space. In the strict asymptotic limit in the direction $u \to \infty$ the cycle $\tau_2$ shrinks to zero and a tower of $\alpha'$ corrections that were neglected here become relevant.}
\subsection{The torus}

The scalings observed in the fibred case also appear in the torus.
As before, assume that the volume is fixed and assume, without loss of generality, that $\tau_1 \ll \tau_2,\, \tau_3$.
The BHP conjecture then states that the dominant contribution to the potential is $\delta V_{\rm KK} (\tau_1)$ (cf.~\eqref{eq:KK potential}).
We want to find a basis $\lbrace\chi\, , \, \chi_1\, , \chi_2 \rbrace$ comprised of the fixed volume modulus ($\chi$), a direction that gets dynamical by the potential ($\chi_1$), and a further direction that remains flat at this order of approximation and whose dynamics we neglect ($\chi_2$).
A straightforward exercise shows that the appropriate change of basis is given by
\begin{equation}
    \chi=M \bar{\chi} \, , \qquad M=
    \begin{pmatrix}
    \frac{1}{\sqrt{3}} & \frac{1}{\sqrt{3}}  & \frac{1}{\sqrt{3}}  \\
    \sqrt{\frac{2}{3}} & -\frac{1}{\sqrt{6}} & -\frac{1}{\sqrt{6}} \\
    0 & \frac{1}{\sqrt{2}} & -\frac{1}{\sqrt{2}}
    \end{pmatrix} \, ,
\end{equation}
where the basis $\lbrace \bar{\chi}_i \rbrace$ is defined via $\tau_{i+1}=\exp(\sqrt{2}\bar{\chi}_i)$.
The potential thus reads, ignoring $\chi$ dependence:
\begin{equation}
    \delta V_{\rm{KK}}\sim e^{-\frac{4}{\sqrt{3}}\chi_1} \, ,
\end{equation}
and so the dynamics of $\chi_1$ is as in the fibred case.

We can thus study the dynamics of branes wrapping the different cycles.
The tension of a D3 wrapping $\t_1$ inherits the following time dependence:
\begin{equation}
    \mu \sim \exp \lr{-\frac{2}{\sqrt{3}}\chi_1}\sim \frac{t_0}{t} \, ,
\end{equation}
while the tension of an NS5 wrapping $\tau_1$ instead grows like $t$.

The time-dependence of the tension of a D3 wrapping any other of the two cycles is not as immediate because it depends on $\chi_2$, which is a flat direction at this level of approximation.
Neglecting the dynamics in this direction (a good approximation due to the combination of Hubble friction and the assumed mass hierarchy), the scalings are
\begin{equation}
    \mu(t) \sim \exp \lr{\frac{1}{\sqrt{3}}\chi_1} \sim \lr{\frac{t}{t_0}}^{1/2}\, ,
\end{equation}
while the tension of an NS5 wrapping the dual cycle instead decreases like $t^{-1/2}$.

We thus conclude that the scalings $\mu \sim t^{\pm \frac{1}{2}}$ and $\mu \sim t^{\pm 1}$ are interesting benchmarks for phenomenological studies of strings with varying tension.\footnote{That the powers are related by a factor of 2 is expected from fixing these particular directions in moduli space.
If $\mu_1 \sim \t_1\sim t^{a_1}$ and $\mu_2 \sim \t_2 \sim t^{a_2}$ one can relate in this direction $\t_1 \sim 1/\t_2\t_3$ (because the volume is fixed) and $\t_2 \sim \t_3$, which is ensured because $\t_2=\t_3$ in the fibre case and because this direction is fixed in the torus case.
It follows that $a_1=-2a_2$.}
This adds to the other known cases of volume modulus kination, with $\mu \sim t^{-1}$ (degenerate with some of the wrapped D3s here) and volume modulus tracker $\mu \sim t^{-2/3}$.
In the remaining of the paper we take a first step in studying their phenomenology.

\section{Gravitational Wave Spectrum}

\begin{figure}[t]
    \centering
    \includegraphics[width=0.9\linewidth]{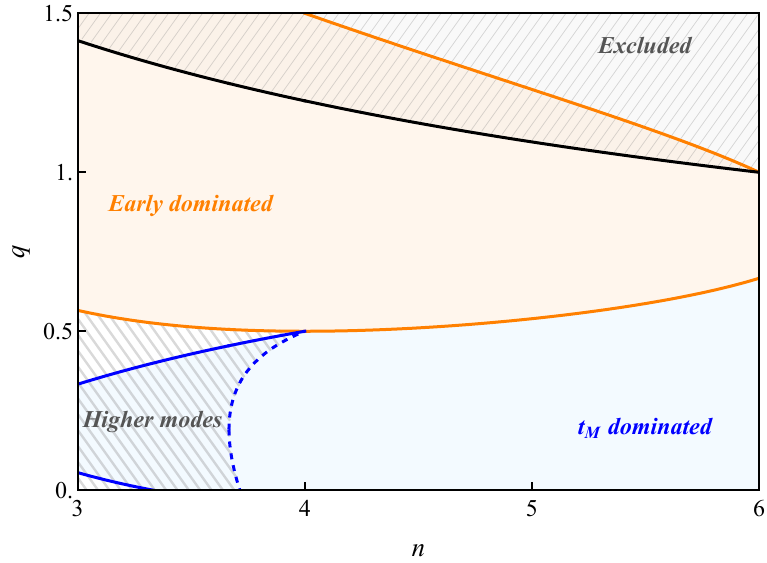}
    \caption{Parameter space ($n,q$), where $G\mu \sim t^{-q}$ and $H=2/(nt)$.
    The main text computes spectral indices of the GW background from scaling networks of cosmic strings in these backgrounds.
    The orange (undashed) region is scenario 1, with spectral index computed in Eq.~\eqref{eq:OGW}.
    The blue (undashed) region is scenario 2, with spectral index computed in Eq.~\eqref{eq:1GW}.
    Dashed regions are either excluded by the Swampland bound in Eq.~\eqref{eq:b3} (right) or have a spectral index of $-1/3$, dominated by higher modes (scenario 3).}
    \label{fig:pn}
\end{figure}

One of the most interesting aspects about scaling configurations of cosmic strings is that their associated GW spectrum carries information about the physics of the early Universe.
As could be expected, variations in the tension of these objects affect the gravitational wave spectrum $\Omega_{\rm{GW}}(f)\equiv \frac{1}{\rho_c}\frac{d \rho_{\rm{GW}}}{d \log f}$, which is expressed in units of the critical energy density of the universe $\rho_c$.
In this last section we discuss some general aspects of the GW spectrum for cosmic strings with varying tension assuming a background with $G\mu (t) =G\mu_0 (t_0/t)^{q}$ and constant equation of state so that the Hubble scale satisfies $H=2/(nt)$ throughout.
We identify the loops that source the largest contribution to the GW spectrum at a given frequency and fully classify the phenomenology in three different scenarios:
\begin{itemize}
    \item \textbf{Scenario 1.}
    The loops radiate the largest contribution to the spectrum at their formation time.
    This is an effect that only occurs when the tension decreases.
    \item \textbf{Scenario 2.}
    The loops radiate the largest contribution to the spectrum when they have radiated half of their energy.
    \item \textbf{Scenario 3.}
    The spectrum is dominated by higher modes of loops sourcing past this time.
\end{itemize}
The regions of $q$ and $n$ where each scenario applies is shown in Fig.~\ref{fig:pn}, in the region of parameter space that is not ruled out by the arguments in the first section.
In what follows we describe the computations that lead to this classification and calculate the resulting spectral indices as a function of $q$ and $n$.
In addition, we discuss the bound that Eq.~\eqref{eq:b3} yields for the spectral index.

\subsection{Qualitative aspects}

We now sketch the steps required in order to compute the GW spectrum for a scaling cosmic string network in the backgrounds of interest.
We assume that the scalings of $G\mu$ and $H$ are as such for times $t_1<\tilde{t}<t_\Delta$, and we only focus on GWs sourced during these times in this paper.
The interested reader is referred to Ref.~\cite{Ghoshal:2025tlk} for details and computations for $\tilde{t}>t_\Delta$.

The gravitational wave spectrum arising\footnote{Here we study the usual spectrum arising from cusps in the string, which gives rise to GWs of typical frequencies related to the length of the loop. Calculations in string theory~\cite{Amati:1999fv} suggest the existence of another decay channel whose emission spectrum peaks at high frequencies, of order the string scale. A study of the phenomenological challenges and opportunities of this channel is left for future work.
We similarly do not discuss other irreducible contributions to the GW background sourced by the strings, such as isocurvature~\cite{Lozanov:2023aez} or non-gaussianities. Apart from resonant effects during fast transitions (see e.g.~\cite{Domenech:2020ssp} in the context of isocurvature), these are suppressed by additional powers of $G\mu$.} from a loop distribution with formation rate $dn/dt_i$ and sourcing GWs at a time $\tilde{t}$ with power $dE/dt$ is given by
\begin{equation}\label{eq:master-gw}
    h^2\Omega_{\rm{GW}} (f)= \frac{ \, f}{\rho_c} \int_{t_1}^{t_\Delta}{d \tilde{t} 
    \left(\frac{a(\tilde{t})}{a_0}\right)^4
    \frac{dE}{d\tilde{t}}
    \left( \frac{a(t_i)}{a(\tilde{t})}\right)^3
    \frac{dn}{dt_i}
    \frac{dt_i}{df}} \, ,
\end{equation}
which redshifts the initial loop population at a time $t_i$ to the emission time $\tilde{t}$ and the sourced energy density to today.
Further, the integrand should be supplemented with Heaviside functions which take into account e.g. the formation time of the network.

The quantity $dn/dt_i$ can be computed from the scaling properties of the network to be $\sim (t_f/t_i)^4$ for some fiducial time $t_f$, and we assume that the emission power is $dE/dt \sim G\mu(t)^2$.
Further assuming that the loops radiate in frequencies corresponding to their vibrational modes, labeled by $k\in \mathbb{Z_+}$, we relate the emission frequency $f_{\rm em}$ to the length of the loop at the emission time $\ell (\tilde{t})$ via $f_{\rm em}=2 k/ \ell(\tilde{t})=f k(a(\tilde{t})/a_0)$.
Let us focus on the effects of the first mode, postponing the discussion of higher modes to later in the section.
We have:
\begin{equation}
    \frac{dt_i}{df}=\frac{dt_i}{d\ell} \frac{d \ell}{df}=-\frac{a(t)}{a_0}\frac{2}{f^2} \frac{dt_i}{d\ell} \, ,
\end{equation}
and the remaining unknown quantity $dt_i/d\ell$ can be computed by relating the creation time and the length at the time of emission via the evolution equation of the loop length, which can be determined itself by the worldsheet equations of motion.
The solution for the evolution equation for $\ell (t)$ is (for $q \neq 2/3$)~\cite{Ghoshal:2025tlk}:
\begin{equation}
    \ell(t)=\ell_i \left( \frac{t}{t_i}\right)^{q/2}
    +\frac{\Gamma G\mu_f}{3q/2-1} \left( \frac{t_f}{t} \right)^{q-1}t_f \, ,
\end{equation}
with $G\mu_f$ the dimensionless tension at a time $t_f$ and $\ell_i$ a constant determined by the initial conditions.
Ref.~\cite{Ghoshal:2025tlk} follows ``Model 1'' in~\cite{LISACosmologyWorkingGroup:2022jok} in assuming this initial condition to be $\ell(t_i)=\alpha t_i$, with $\alpha$ an order one number which may depend on the equation of state of the background or the intercommutation probability of the network.
Thus,
\begin{equation}
    \ell_i=\left(\alpha -\frac{\Gamma G\mu_f}{3q/2-1} \left( \frac{t_f}{t_i}\right)^{q}\right)t_i \simeq \alpha t_i \, ,
\end{equation}
where in what follows we will work under the simplifying assumption that $R (t) \equiv \Gamma G\mu(t)/(\alpha(1-3q/2))\ll 1$ - for positive $q$ this is always justified provided the initial tension is $G\mu_f \ll 1$, which is required for a perturbative description of the system.
We can thus work out the formation time of a loop that at a time $t$ sourced GWs that today have a frequency $f$:
\begin{equation}\label{eq:ti}
    \left( \frac{t_i}{t_f}\right)^{1-q/2} = 
    \left( \frac{t_f}{t}\right)^{q/2}
    \left[ \frac{a(t)}{a(t_f)}\frac{f_f}{f}+R_f\left(\frac{t_f}{t} \right)^{q-1}
    \right] \, ,
\end{equation}
where $f_f$ is a fiducial frequency related to $t_f$ and $R_f = R(t_f)$.

Eq.~\eqref{eq:ti}, at fixed $f$, distinguishes two regimes: either $t_i$ is dominated by the scale factor-dependent contribution or by the term proportional to $R_f$.
The case $q>1-2/n$ where the first term always dominates is discussed later.
For $q<1-2/n$ instead, the second term is subdominant until a time $t_M(f)$ where both contributions become equal, namely
\begin{equation}
    \frac{a(t_M)}{a(t_f)}\frac{f_f}{f}=R_f\left(\frac{t_f}{t_M} \right)^{q-1} \, ,
\end{equation}
which defines $t_M$ implicitly.
It is possible to show that
\begin{equation}
    l(t_M) =\frac{l_i}{2} \left(\frac{t_i}{t_M}\right)^{q/2} \, ,
\end{equation}
which implies that $t_M$ is the time at which the loops have lost half of their initial energy.

The computation then inserts the relevant quantities described above into the integral in Eq.~\eqref{eq:master-gw}, finding a different power-law dependence on $\tilde{t}$ depending on whether $\tilde{t}<t_M(f)$ or $\tilde{t}>t_M(f)$.
For $\tilde{t}<t_M(f)$, one finds that the overall dependence on $\tilde{t}$ is
\begin{equation}\label{eq:integral}
     \frac{\Omega_{\rm GW}}{\Omega_{\rm GW,\Delta}}=\left(\frac{f}{f_\Delta}\right)^D \int_{t_1}^{t_M (f)} d\tilde{t} \, \left(\frac{\tilde{t}}{t_\Delta}\right)^E \, , \quad D=\frac{6(n-2)}{n(2-q)} \, ,
\end{equation}
with
\begin{equation}
     E=\frac{4}{n}-\frac{5q}{2}+\frac{qn-4}{n(q-2)}\left(\frac{6}{n}+\frac{q}{2}-4 \right)\, .
\end{equation}
All we are interested in the present article is the frequency-dependence of the GW spectrum.
For completeness, however, we quote here that the frequency-independent contribution $h^2\Omega_{\rm GW,\Delta}$ can be estimated in terms of redshift $z_\Delta$ and the microscopic parameters of the string at the relevant time, $G\mu(t_\Delta)$ and $P$, as
\begin{equation}\label{eq:univ-amplitude}
    \Omega_{\rm GW,\Delta} = K_\Delta 
    \frac{\lr{G\mu(t_\Delta)}^2}{P^{2(1-\beta)+\chi}}
    \lr{\frac{t_0}{t_\Delta z_\Delta^2}}^2\, ,
\end{equation}
where $\beta$ and $\chi$ are parameters capturing the different $P$-dependences of cosmic string physics reported in the literature, as discussed in~\cite{Ghoshal:2025tlk}.
The constant $K_\Delta$ depends on simulation-inferred quantities which do not play any role in the discussion and correspond to a minor scaling of the overall amplitude, though we refer the reader to~\cite{Ghoshal:2025tlk} for definitions.

\subsection{The different scenarios}

The physics is qualitatively different depending on the sign of $E+1$ in Eq.~\eqref{eq:integral} and this allows us to distinguish the different scenarios.

\noindent \textbf{Scenario 1.}

Let us discuss the first case, $E<-1$, where the GW spectrum at frequency $f$ is dominated by loops that sourced GWs at the appropriate frequency bin during their formation time, namely $t_i(t_{\rm min}(f),f)=t_{\rm min}(f)$.
This occurs for $q_{-}(n)<q <q_+ (n)$, where
\begin{equation}\label{eq:qpm}
    2n \, q_\pm (n)=4+n\pm\sqrt{-3n^2 +24n-32} \, .
\end{equation}
If so, from Eq.~\eqref{eq:ti}, we see that
\begin{equation}
    \frac{a(t_{\rm min})}{a_0} \frac{f_0}{f}=\frac{t_{\rm min}}{t_0} \left(1- R_0\left( \frac{t_0}{t_{\rm min}}\right)^q \right) 
    \simeq \frac{t_{\rm min}}{t_0}\, .
\end{equation}
Performing the integral and focusing on the contribution from $t_{\rm min}(f)$, this gives rise to \begin{equation}\label{eq:OGW}
    \Omega_{\rm{GW}} = \Omega_{\rm{GW,0}} \lr{\frac{f}{f_0}}^A\, , \quad A\equiv 2 \left[1+\frac{qn-2}{n-2}\right]\, .
\end{equation}
The region where Eq.~\eqref{eq:OGW} holds is plotted in orange in Fig.~\ref{fig:pn}, with the orange solid lines corresponding to $q_{\pm}(n)$ as per Eq.~\eqref{eq:qpm}.
The region at high $q$ is excluded by the decompactification bounds in Eq.~\eqref{eq:b3}, denoted by a solid black line.

\noindent \textbf{Scenario 2.}

Let us now study the regime where the GW spectrum is dominated by $t_M$.
This requires both $E>-1$ and $q<1-2/n$.
If $q>1-2/n$ with $E>-1$, $t_M(f)$ does not exist and the integral is dominated by very late times.
This case belongs to the third scenario and the boundary $q=1-2/n$ is the uppermost solid blue line in Fig.~\ref{fig:pn}.

In this scenario, the integral should be performed and evaluated now at $t_M(f)$, giving rise to $\Omega_{\rm GW} \sim f^B$, with
\begin{equation}\label{eq:1GW}
    B = 2 \left[\frac{8-10q-2n+2qn+nq^2}{(2-q)(2+n(q-1))}\right]\, .
\end{equation}
One can similarly evaluate the integral with the ``late time" expression for $t_i$, namely the second term in Eq.~\eqref{eq:ti}, and show that the integral is dominated by $t_M(f)$ for $2 nq> 10/n-3$.
This is the lowermost solid blue line in Fig.~\ref{fig:pn}.
The integral is otherwise dominated at later times, which belong to the third scenario.

\noindent \textbf{Scenario 3.}

In the cases that do not belong to the above the spectral index is evaluated at $t_\Delta$ where the description breaks down, either because the equation of state or the rate of variation of the tension change.\footnote{This also the case when $t_M(f)<t_\Delta$, which can be used to compute values of $f$ where the spectral index changes~\cite{Ghoshal:2025tlk}.}
To fully compute the contribution to the GW spectrum from the loops formed during the epoch $\tilde{t}<t_\Delta$ that we are currently discussing one should solve for the evolution of $\ell(t)$ in the new background and proceed as above, see~\cite{Ghoshal:2025tlk} for details, but this remains beyond the scope of this work.
Focusing on the epoch $\tilde{t}<t_\Delta$ within the white region in Fig.~\ref{fig:pn}, the integral in Eq.~\eqref{eq:master-gw} evaluates at $t_\Delta$ and so $\Omega_{\rm GW} \sim f^{-1}$.
In this case higher modes play an important role.
As discussed in~\cite{Blasi:2020wpy}, when the first mode predicts a spectral index $f^{C}$, $C<-1/3$, higher modes tend to modify it to $-1/3$. This physics occurs not only within the white region, since there are values of $n$ and $q$ where $B\leq-1/3$ within the blue region in Fig.~\ref{fig:pn}.
One can compute that, within the blue region, $B\leq -1/3$ for $n<2(31q-26)/(5q^2+15q-14)$ (dashed blue in the figure), and so the spectral index in this region is not given by Eq.~\eqref{eq:1GW} but by $B=-1/3$.

\subsection{Bounds on the spectral index}

In combination with the results of this section, the Swampland-inspired bounds on the time dependence of the tension can be used to constrain the spectral index of the corresponding GW signal.
In the case of decompactification limits, and assuming a scaling solution for the background cosmology, one can insert \eqref{eq:b3} into \eqref{eq:OGW} to bound the spectral index in the case where the emission is dominated by early times (scenario 1), obtaining
\begin{equation}\label{eq:gwb}
    A \leq 2\left[ 1+\frac{\sqrt{6n}-2}{n-2}\right] .
\end{equation}
If, instead, the GW emission is dominated at a time $t_M$, (corresponding to scenario 2), the expression for the spectral index \eqref{eq:1GW} is a monotonically increasing function of $q$ for fixed $n$, and coincides with \eqref{eq:OGW} when $q=q_-$ (\emph{i.e.} at the boundary between the blue and orange regions in Fig. \ref{fig:pn}). Therefore, the spectral index is still bounded by \eqref{eq:gwb}. Moreover, for the white region of Figure \ref{fig:pn} where scenario 3 is realised, the spectral index is negative. We conclude that the bound \eqref{eq:gwb} holds in the entire parameter space, given the assumptions we have just stated.

Although the bound \eqref{eq:gwb} would appear to become less stringent for small values of $n$ (\emph{e.g.} $n=3,4$ relevant for matter/radiation trackers), we recall that \eqref{eq:b3} can only be saturated in the case of field-dominated scaling solutions. The latter require a very shallow scalar potential, and are typically unstable, as remarked in previous sections. It would be a very interesting avenue of future work to try to understand whether there exist any physically relevant cases where the upper bound of \eqref{eq:gwb} can be reached. Indeed, this does not happen in any of the examples discussed in the literature, which only reach up to $q=1$. This corresponds to $A=4$, which is independent of the equation of state of the background.
It is very interesting that there are examples in string theory that reach the line $q=1$ for $n=3$, 4 (wrapped branes in tracker regimes, as above) and 6 (volume modulus kination)~\cite{Conlon:2024uob}.

\section{Conclusions}

In string compactifications, cosmic strings can arise both from fundamental (F1) strings as well as $p$-branes wrapping a compact $p-1$ cycle. In the latter case, their tension is controlled by the volume of the individual cycle they wrap, and can naturally acquire a time-dependence even if the overall volume and the dilaton are stabilised. A first result of this work are bounds on the rate of variation for the tension of \emph{any} such effective string, Eqs. \eqref{eq:mubound}-\eqref{eq:b1}, assuming the validity of certain Swampland conjectures (such as the ESC).

Phenomenologically, periods of decreasing tension in the early universe are very interesting in that they give rise to GW spectra with spectral indices that are larger than the constant tension case, boosting the amplitude at high frequencies.
Building on \cite{Ghoshal:2025tlk}, we derived analytic expressions for the spectral index expected from a scaling network of such objects, in the case of a power-law background cosmology. The Swampland-inspired bounds then allowed us to derive the universal constraint ~\eqref{eq:gwb} on their spectral index, assuming the dynamical moduli are exploring a decompactification (rather than tensionless string) limit.
It would be interesting to understand whether the connection to Swampland constraints could be taken further, perhaps linking the tension of the strings to other properties of the compactification \cite{Lanza:2020qmt,Lanza:2021udy,Grieco:2025bjy}.

On the other hand, the precise value of the spectral index in a given case relies on the dependence of the string tension on the various moduli, and also from how the latter evolve with time. Therefore, any realistic calculation requires knowledge of the potential for the moduli, and eventually of a scenario for moduli stabilisation (\emph{i.e.} the existence of a late-time minimum). For this reason, we specialised to the phenomenologically motivated case of type IIB compactifications, with D3 and NS5 branes wrapping $2-$ and $4-$cycles in fibred or toroidal manifolds. In both cases, we found that the string tensions follow the general scalings $\mu \sim t^{\pm \frac{1}{2}}$ or $\mu \sim t^{\pm 1}$, depending on which cycle the branes wrap. The case $\mu \sim t^{-1}$ is particularly interesting from a phenomenological perspective, as it implies a GW spectral index that is independent of the background equation of state.

Our work significantly expands previous discussions~\cite{Ghoshal:2025tlk} in studying the GW phenomenology of scaling networks of cosmic strings with varying tension in all well motivated scenarios.
Using the Swampland bounds to identify a well motivated region in the $(n,q)$ plane, we have identified three scenarios and computed their associated spectral indices.
These are the differently colored regions of Fig.~\ref{fig:pn}.

In the present paper, we have focused on the spectral index of the GW spectrum without mention of the overall amplitude or the relevant frequency ranges.

These strongly depend on the detailed cosmological history of the universe, which can vary between models.
In particular, GWs emitted in a pre-BBN tracker phase may be strongly diluted by any subsequent period of matter domination, 
ubiquitous in models where reheating is driven by decaying moduli \cite{Cicoli:2012aq,Cicoli:2022fzy}.
A detailed study of the amplitude and frequency ranges where Eq.~\eqref{eq:OGW} applies has been carried out in this context in~\cite{Ghoshal:2025tlk}. Applying such methods to a concrete embedding of our wrapped-brane scenarios is a promising avenue for further investigation. 

\appendix

\section{Acknowledgements}
We are grateful to José Calderón Infante, Alberto Castellano, Michele Cicoli, Joe Conlon, Thomas Grimm, Damian van de Heisteeg, Álvaro Herráez, Antonio Iovino, Stefano Lanza, Francisco G. Pedro, Fernando Quevedo, Marco Scalisi and Flavio Tonioni for discussions and comments on the draft. FR is supported from a junior postdoctoral fellowship of the Fonds Wetenschappelijk Onderzoek (FWO), project number 12A1Q25N.
GV is supported by a Research Fellowship from Gonville \& Caius College, Cambridge.
This work has been partially supported by STFC consolidated grant ST/X000664/1.
The work of LB, FR and GV contributes to the COST Action COSMIC WISPers CA21106, supported by COST (European Cooperation in Science and Technology).

\bibliographystyle{jhep}
\bibliography{biblio}

\end{document}